\documentclass[conference,10pt]{IEEEtran}
\IEEEoverridecommandlockouts
\usepackage{cite}
\usepackage[colorlinks, linkcolor=magenta, anchorcolor=green, citecolor=blue]{hyperref}
\usepackage{amsmath,amssymb,amsfonts}
\usepackage[numbers,sort&compress]{natbib}
\usepackage{algorithmic,algorithm}
\usepackage{fancyhdr}
\usepackage{graphicx}
\usepackage{textcomp}
\usepackage{xcolor}
\def\BibTeX{{\rm B\kern-.05em{\sc i\kern-.025em b}\kern-.08em
    T\kern-.1667em\lower.7ex\hbox{E}\kern-.125emX}}
\usepackage{multirow}
\usepackage{makecell}

\usepackage[top=0.75in,bottom=0.80in,left=0.58in,right=0.58in]{geometry}
\iftrue
\fi

\begin{document}

\title{\huge Towards Affordable, Adaptive and Automatic GNN Training on CPU-GPU Heterogeneous Platforms\\
% {
% \footnotesize \textsuperscript{*}Note: Sub-titles are not captured in Xplore and
\thanks{
This work is supported in part by the National Natural Science Foundation of China (Grant No. 62072019), the Beijing Natural Science Foundation (Grant No. L243031), the National Key R\&D Program of China (Grant No. 2023YFB4503704 and 2024YFB4505601).
Corresponding author is \textit{Jianlei Yang}, Email: \url{jianlei@buaa.edu.cn}}
}

\author{
    \IEEEauthorblockN{
        Tong Qiao\IEEEauthorrefmark{2},
        Ao Zhou\IEEEauthorrefmark{2},
        Yingjie Qi\IEEEauthorrefmark{2},
        Yiou Wang\IEEEauthorrefmark{2},
        Han Wan\IEEEauthorrefmark{2},
        Jianlei Yang\IEEEauthorrefmark{2}\IEEEauthorrefmark{3} and
        Chunming Hu\IEEEauthorrefmark{4}
    }
    \IEEEauthorblockA{
        \IEEEauthorrefmark{2}School of Computer Science and Engineering, Beihang University, China \\
        \IEEEauthorrefmark{3}Qingdao Research Institute, Beihang University, Qingdao, China \\
        \IEEEauthorrefmark{4}School of Software, Beihang University, China
    }
}

\maketitle
\thispagestyle{empty}
\pagestyle{empty}

\bstctlcite{IEEEexample:BSTcontrol}

\begin{abstract}

Graph Neural Networks (GNNs) have been widely adopted due to their strong performance. 
However, GNN training often relies on expensive, high-performance computing platforms, limiting accessibility for many tasks. 
Profiling of representative GNN workloads indicates that substantial efficiency gains are possible on resource-constrained devices by fully exploiting available resources. 
This paper introduces $\textrm{A}^{\textsubscript{3}}$GNN, a framework for \underline{A}ffordable, \underline{A}daptive, and \underline{A}utomatic GNN training on heterogeneous CPU-GPU platforms. 
It improves resource usage through locality-aware sampling and fine-grained parallelism scheduling. 
Moreover, it leverages reinforcement learning to explore the design space and achieve pareto-optimal trade-offs among throughput, memory footprint, and accuracy. 
Experiments show that $\textrm{A}^{\textsubscript{3}}$GNN can bridge the performance gap, allowing seven Nvidia 2080Ti GPUs to outperform two A100 GPUs by up to 1.8$\times$ in throughput with minimal accuracy loss.

\end{abstract}


\vspace{-6pt}
\section{Introduction}

Graph Neural Networks (GNNs) have demonstrated significant success across diverse domains, such as recommendation systems~\cite{yang2023dgrec}, link predictions~\cite{wang2024efficient}, and knowledge graph analysis~\cite{bi2023bridged}, yielding remarkable achievement.
Due to their intrinsic ability to capture structural information, GNNs play a pivotal role in handling large-scale graph data~\cite{yang2023betty, liu2023bgl}.
As the volume of real-world graph explodes, a widening gap is emerging between computing capabilities and GNN training requirements~\cite{yingjie2023arch}.
Effectively harnessing the power of high-performance GPUs is crucial for bridging this computational gap in large-scale GNN training.

Many efforts have been focused on optimizing GNN training performance on CPU-GPU heterogeneous platforms, specifically addressing three key aspects of performance: throughput, peak GPU memory footprint, and test accuracy.
To enhance system throughput, Quiver~\cite{quiver2023} and DSP~\cite{cai2023dsp} execute sampling on GPUs to reduce sampling overhead, while BGL~\cite{liu2023bgl} utilize idle GPU memory resources as cache to mitigate feature retrieval costs.
For GPU memory optimization, PaGraph~\cite{lin2020pagraph} executes input layer of GNNs on CPU, alleviating the memory demands from activations on the GPU.
EXACT~\cite{liu2021exact} designs a novel data structure for further activation compression, aiming to minimize peak memory footprint.

\begin{figure}[t]
    \centering
    \includegraphics[width=1\linewidth]{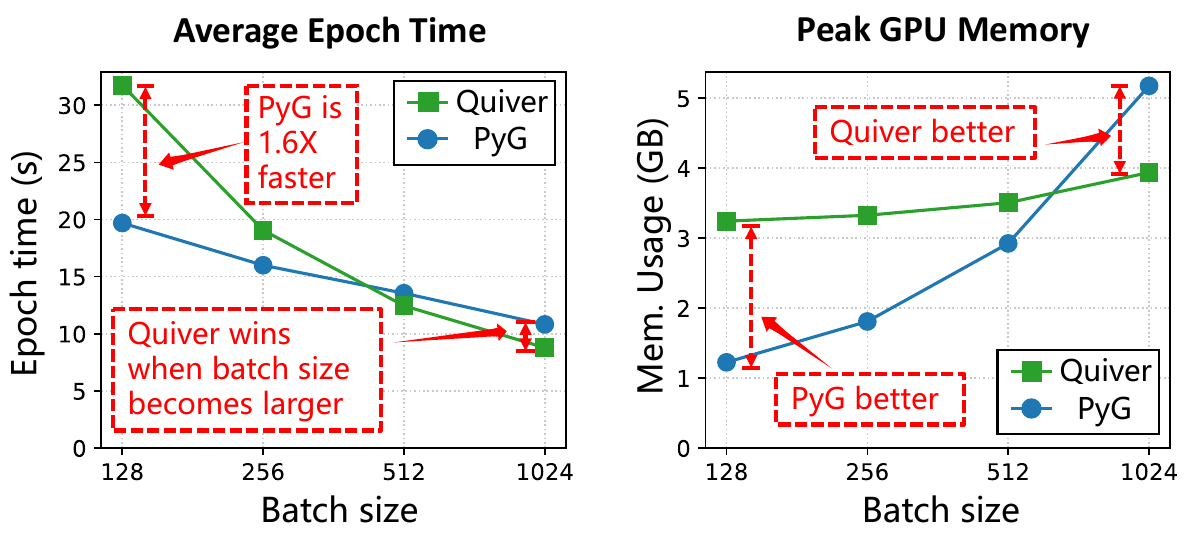}
    \vspace{-18pt}
    \caption{Training performance profiling of PyG and Quiver.}
    \label{fig:profile}
\end{figure}

However, existing approaches are insufficient to optimize GNN training on resource-constrained platforms (for example, desktop platforms with Nvidia 2080Ti GPU)~\cite{ao2023hardware}.
While many works focus on accelerating GNN training on high-performance computing platforms, the dependence on costly devices limits their accessibility.
Thus, a key challenge is enabling equitable research and affordable GNN training.
Furthermore, the efficacies of existing solutions vary across different configuration settings, suggesting that no single approach is optimal across all scenarios.
As illustrated in Fig.~\ref{fig:profile}, PyG~\cite{Fey/Lenssen/2019} excels with smaller batch sizes, while Quiver~\cite{quiver2023} outperforms with larger ones.
Finally, the reliance on manual parameter tuning, a time-consuming and sub-optimal process, further highlights the need for efficient automatic training optimization.

In this paper, we present $\textrm {A}^{\textsubscript{3}}$GNN, a novel GNN training framework designed for heterogeneous compute platforms. 
Through adaptive and automatic resource management and parallelism scheduling, $\textrm {A}^{\textsubscript{3}}$GNN can achieve significant training performance improvements under an affordable budget.
Compared to existing GNN training approaches, the key distinction of our framework lies in its hardware-aware optimizations and auto-tuning capabilities.
Our contributions are summarized as follows:
\begin{itemize}
    \item \textbf{Affordable and efficient locality-aware sampling.} We develop a cache-aware sampling strategy to enhance data locality by prioritizing the selection of already cached nodes.
    \item \textbf{Adaptive multi-level parallelism scheduling.} We design a multi-level parallel pipelining mechanism tailored for CPU-GPU heterogeneous platforms, which adaptively adjusts parallelism to maximize performance.
    \item \textbf{Automatic configuration settings fine-tuning.} We employ reinforcement learning based Pareto front exploration for fine-grained auto-tuning of training configurations.
\end{itemize}
\section{Preliminaries \& Motivations}

\subsection{Distributed GNN Training with Multi-GPUs}

Algo.~\ref{distgnnalgo} outlines the distributed GNN training process on multi-GPUs platforms.
To mitigate communication overhead arising from frequent remote data requests, each GPU manages its own partition of the subgraph data.
Firstly, the dataloader iteratively yields batch indices from a subgraph $Gs_i$ iteratively (line 5-7). 
A batch is then generated according to these indices, requiring communication with memory or other GPUs when necessary (line 9-10).
The GPUs subsequently train the GNN model using the generated batch and synchronize model parameters (line 12-14)~\cite{ao2021bip}. 
Finally, the trained model $M_{trained}$ is returned for further processing.

\begin{algorithm}[t]
\small % 使用更小的字体
% \setstretch{0.9} % 减少行间距
\caption{Distributed GNN Training with Multi-GPUs.}
\renewcommand{\algorithmicrequire}{\textbf{Input:}}
\renewcommand{\algorithmicensure}{\textbf{Output:}}
\begin{algorithmic}[1]
\label{distgnnalgo}
\REQUIRE Initial GNN model $M_{init}$, graph data $G(V,E)$, number of GPUs $u$.
\ENSURE Trained GNN model $M_{trained}$.
\STATE \textit{--- Optional ---}
\STATE \{$Gs_1,Gs_2,..Gs_u$\}=graph\_partition($G$) 
\FOR {$i < u$}
    \STATE \textit{--- Sampling ---}
    \STATE Loader = new(Dataloader($Gs_i$))
    \WHILE{$index$ = Loader.get()}
        \STATE $local\_index$ = reindex(index, $G$)
        \STATE \textit{--- Batch Generating ---}
        \STATE communicate()
        \STATE $batch$ = feature\_retrieve($local\_index$, $Gs_i$)
        \STATE \textit{--- Training ---}
        \STATE forward($batch$, $M$)
        \STATE backward()
        \STATE sync()
    \ENDWHILE
\ENDFOR
\RETURN $M_{trained}$
\end{algorithmic}
\end{algorithm}

\subsection{Related Works and Motivations}

Drawing on prior research, we offer $3$ key observations that inspire affordable, adaptive and automatic GNNs training on CPU-GPU heterogeneous computing platforms.

\textbf{Observation 1: Current GNN training optimizations are designed primarily for high-performance computing platforms.}
As real-world graphs continue to grow in scale, GNN training calls for more effective strategies for the increasing training workloads.
To tackle the performance bottlenecks of GNN training on large-scale graphs, recent works have concentrated on communication optimizations, sampling techniques, and pipeline parallelism in distributed GNN systems.
Taking communication optimization methods as an example, ParallelGCN~\cite{demirci2022scalable} utilizes graph partitioning with peer-to-peer (P2P) protocols to reduce data transfer volume.
BNS-GCN~\cite{wan2022bns} explores distributed sampling primitives, optimizing memory footprint and time overhead while accepting some accuracy trade-offs.
Other works, such as BGL~\cite{liu2023bgl}, and FlexGraph~\cite{wang2021flexgraph}, attempt to overlap communication and computation for improved system throughput.
Additionally, some approaches, like DeepGalois~\cite{hoang2021efficient} and PaGraph~\cite{lin2020pagraph}, perform partial aggregation prior to communication to reduce network overhead.
Similarly, sampling and pipeline optimization strategies also heavily rely on high-end hardware capabilities.
Advanced sampling techniques like GNNLab~\cite{yang2022gnnlab} require substantial memory bandwidth for efficient neighbor sampling operations, due to their frequent data exchange between CPUs and GPUs.
Pipeline optimization frameworks such as TurboGNN~\cite{wu2023turbognn}, Blad~\cite{fu2023blad} and Legion~\cite{sun2023legion} depend on high-speed interconnects and large memory capacity to maintain multiple concurrent pipeline stages effectively.
However, existing optimizations heavily depend on specialized hardware, with frameworks like DSP~\cite{cai2023dsp} and BGL~\cite{liu2023bgl} necessitating expensive high-performance GPUs with NVLink and NVSwitch for efficient data transfer.
On resource-constrained platforms, such as desktop platforms with 2080Ti GPUs, most existing approaches either fail to support these optimization strategies or exhibit poor training performance due to hardware limitations.
This hardware dependency severely limits the applicability of existing optimizations on resource-constrained platforms, necessitating adaptive strategies that can coordinate multiple optimization techniques based on available resources.

\begin{figure}[t]  % [t] 表示靠顶部，可以改为 [b] 或 [h]
  \centering
  \includegraphics[width=1.05\linewidth]{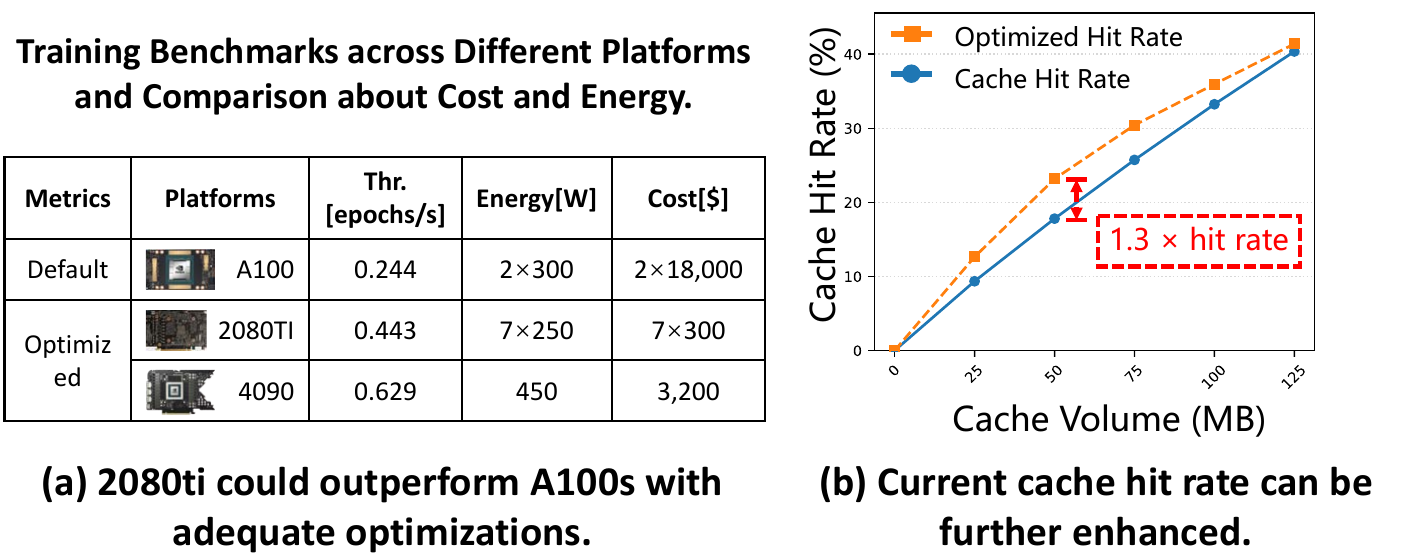}
  \vspace{-18pt}
  \caption{Profiling GNNs training tasks on different datasets and platforms.}
  \label{fig:motivation}
\end{figure}

\textit{Motivation 1: Adaptive resource allocation and training strategy coordination can enable resource-constrained devices to outperform high-end systems.}
To validate this hypothesis, we conducted preliminary experiments as shown in Fig.~\ref{fig:motivation}(a).
2080Ti GPUs in our experiments operate without inter-GPU communication, using adaptive parallel scheduling based on hardware resources.
This configuration is labeled as 'optimized'.
As a baseline, we profile the training with PyG's default training settings on A100 GPUs, denoted as "default".
Fig.~\ref{fig:motivation}(a) illustrates how a cluster of seven 2080Ti GPUs can achieve comparable or even superior performance to two A100 GPUs.
The results demonstrate that optimized configuration settings enable cost-effective GNN training with reduced energy consumption, even on less powerful devices.

\textbf{Observation 2: Existing works fail to fully utilize the locality characteristics of graph data.}
As one of the most effective training optimization techniques, prior works including PaGraph~\cite{lin2020pagraph}, TurboGNN~\cite{wu2023turbognn} and DSP~\cite{cai2023dsp} have demonstrated significant performance improvements through feature caching.
PaGraph~\cite{lin2020pagraph} employs a straightforward yet effective approach by introducing the concept of node "hotness", defined by out-degree, to guide the prioritized caching of node features in GPU memory.
A dynamic caching mechanism, leveraging a first-in, first-out (FIFO) replacement policy for cached node features, is introduced in BGL~\cite{liu2023bgl} and GNNavigator~\cite{qiao2024gnnavigator} to achieve higher cache hit rates.
However, these cache-based approaches fail to integrate sampling strategies with caching policies.
They still sample node neighborhoods uniformly without considering cached data availability.
To improve cache utilization, Quiver~\cite{quiver2023} implements clique-based independent caching, enabling each device within a clique to cache a unique subset of feature data.
Legion~\cite{sun2023legion} optimizes distributed caching across multiple devices by introducing a sophisticated unified cache design.
Methods such as Quiver and Legion require substantial hardware resources, leading to significant performance degradation on resource-constrained platforms.
However, on resource-constrained platforms, GNN training already consumes most available GPU memory, making it difficult to allocate sufficient cache space.
Consequently, memory limitations severely restrict cache volume, resulting in diminished acceleration when caching only small subsets of feature data.

\textit{Motivation 2: Coordinated sampling-caching strategies can overcome memory constraints by maximizing cache utility on resource-limited platforms.}
Traditional approaches treat sampling and caching as independent processes, missing critical optimization opportunities.
Our preliminary experiments demonstrate that by biasing sampling toward already-cached nodes, we can achieve 30\% higher cache hit rates (Fig.~\ref{fig:motivation}(b)), significantly amplifying the benefits of limited cache space.
This coordination becomes particularly crucial on resource-constrained platforms where GPU memory limitations severely restrict cache volume. 
Rather than simply caching more data, intelligent sampling-aware caching policies can extract maximum value from minimal cache allocations. 

\textbf{Observation 3: Different configuration settings lead to significant performance variations across platforms.}
It is well understood that no single training configuration setting can achieve optimal performance across all hardware platforms. 
Taking GNN training pipeline optimization as an example, as shown in Algo.~\ref{distgnnalgo}, GNN computation can be decoupled into three main stages: sampling (involving extensive random memory access along graph topology), batch generation (primarily data movement), and training (computation-intensive forward and backward passes).
By carefully analyzing data dependencies between stages and designing pipeline-based parallel strategies that map different phases to separate components, training speed can be significantly improved and system throughput substantially increased.
Several pipeline optimization approaches demonstrate different design choices:
SALIENT~\cite{kaler2022accelerating}, as an exemplary two-stage pipeline work, achieves parallelism between training and batch generation to improve data throughput. 
TurboGNN~\cite{wu2023turbognn}, Legion~\cite{sun2023legion}, and DSP~\cite{cai2023dsp} adopt fixed three-stage pipeline strategies, implementing producer-consumer patterns and degree-guided adaptive shared memory optimizations to enhance pipeline performance.
Recent works~\cite{liu2023bgl, qiao2024gnnavigator} argue for more fine-grained stage partitioning to further optimize training performance.

\textit{Motivation 3: Automated configuration tuning is essential to navigate the complex parameter space of coordinated optimization strategies.} 
For instance, on platforms with strong CPUs but weak GPUs, GPU-based operations become bottlenecks, requiring increased CPU worker allocation to achieve better load balancing.
Conversely, on memory-constrained devices, overly fine-grained pipelines can incur $3$-$4\times$ memory overhead during peak usage, requiring reduced pipeline complexity to prevent memory exhaustion.
The vast configuration space of pipeline scheduling parameters makes manual optimization practically infeasible.
Therefore, this complexity necessitates automated tuning mechanisms that can explore the configuration space efficiently and adapt to both hardware constraints and task-specific requirements in real-time.

These observations collectively highlight a critical gap: 
existing GNN training systems lack adaptive, resource-aware optimization frameworks that can automatically coordinate multiple optimization strategies while dynamically adjusting to diverse hardware environments and task-specific demands. 
This motivates our design of $\textrm {A}^{\textsubscript{3}}$GNN, which addresses all three challenges through intelligent resource coordination, locality-aware sampling optimization, and automated parameter tuning mechanisms.

\vspace{-2pt}
\section{The $\textrm {A}^{\textsubscript{3}}$GNN Framework}
\label{sec:framework}
\begin{figure}[t]
    \centering
    \includegraphics[width=0.96\linewidth]{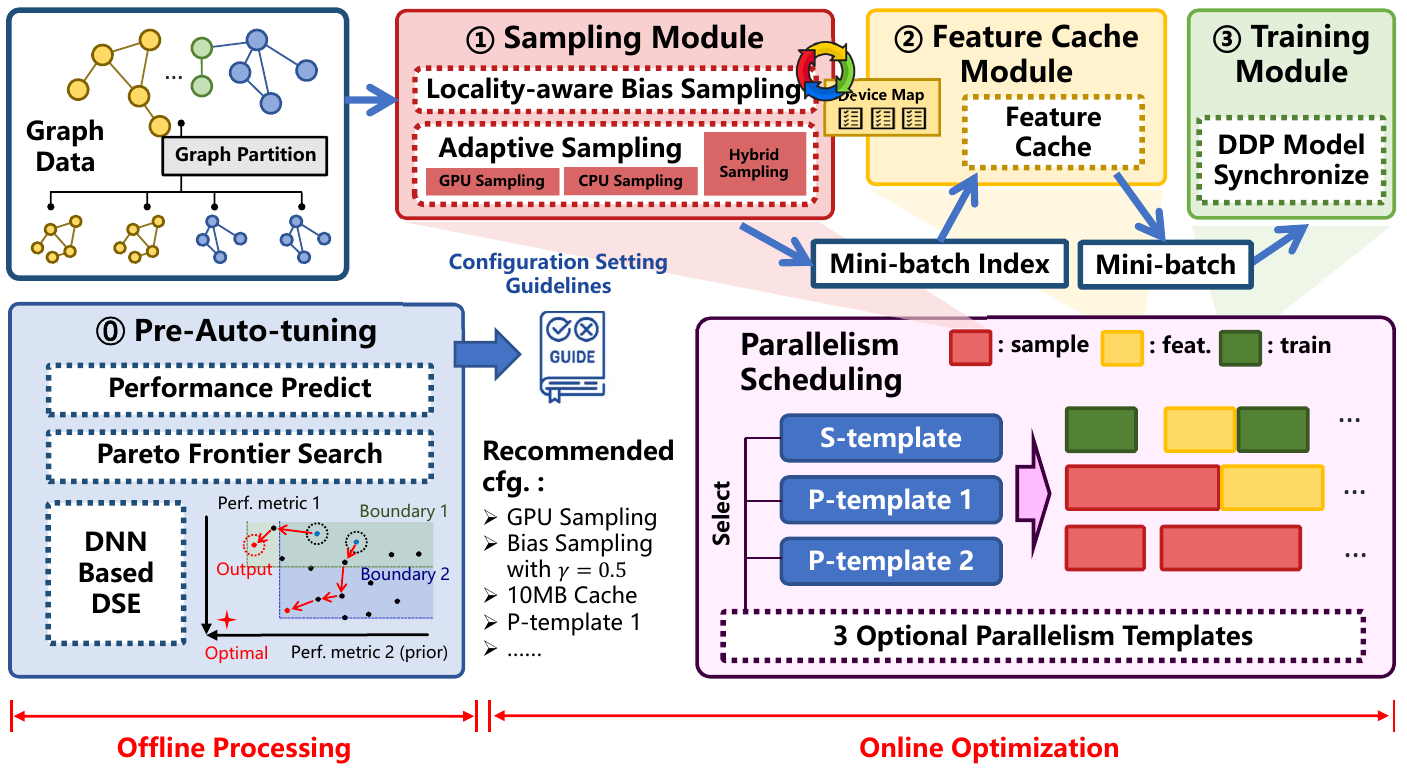}
    \vspace{-8pt}
    \caption{Framework overview of $\textrm {A}^{\textsubscript{3}}$GNN.}
    \label{fig:overview}
    \vspace{-2pt}
\end{figure}

Fig.~\ref{fig:overview} shows the overall workflow of $\textrm {A}^{\textsubscript{3}}$GNN.
The GNN training task is decoupled and mapped onto $3$ modules: 1) sampling module, which incorporates various sampling techniques such as locality-aware biased sampling; 2) feature cache module, which supports diverse caching strategies and maintains a device map for efficient lookup to enable biased sampling; and 3) training module, which is responsible for model training.

$\textrm {A}^{\textsubscript{3}}$GNN achieves automatic, adaptive, and affordable GNN training through two key components: an offline auto-tuning module and an online parallelism scheduling module.
To enable efficient optimization, the auto-tuning module employs a profiling-based approach, training a surrogate model using public datasets from diverse tasks.
This surrogate model predicts GNN training performance across three metrics, thereby avoiding the substantial overhead of ground-truth GNN training profiling and greatly improving optimization efficiency.
The performance metrics mainly include 1) throughput, 2) memory footprint, and 3) test accuracy.
Adaptivity and affordability are achieved by the pre-auto-tuning module, which selects training strategies and configurations based on hardware compute resources, graph data characteristics, and specific task requirements, generating configuration setting guidelines for the GNN training process.
For instance, the decision of whether to execute sampling on GPUs or CPUs is contingent upon the relative GPU/CPU compute capability across hardware platforms.
Therefore, automatically and adaptively balancing the various performance metrics according to specific tasks becomes a complex trade-off.

\vspace{-3pt}
\subsection{Locality-Aware Graph Sampling}
Graph sampling can significantly bottleneck system throughput under various conditions and naive sampling strategies may cause neighborhood explosion.
Consequently, enhancing data locality has become a key concern, particularly by maximizing cache hit rates within a limited cache volume.
To address these challenges, we designed a locality-aware graph sampling strategy.
This strategy prioritizes vertices with cached features, guided by a specified bias rate, to improve data locality.
Experimentally, this locality-aware sampling strategy alone yields up to a $1.30\times$ improvement in system throughput while maintaining acceptable accuracy.

\begin{algorithm}[t]
\small % 使用更小的字体
% \setstretch{0.9} % 减少行间距
\caption{Weighted Reservoir Sampling.}
\renewcommand{\algorithmicrequire}{\textbf{Input:}}
\renewcommand{\algorithmicensure}{\textbf{Output:}}
\begin{algorithmic}[1]
\label{bias-sample}
\REQUIRE $\mathcal{N}(v_i)$ is the neighbor set of $v_i$ , the weight of $u_j$ in $\mathcal{N}(v_i)$ denoted as $w_j$
\ENSURE Sampled subset $R$ with $m$ elements
\FOR {$u_j \in \mathcal{N}(v_i)$}
    \STATE $k_j = \text{pow}(\text{rand}(0, 1), 1/w_j)$
    \IF {$j < m$}
        \STATE Add $(u_j, k_j)$ to $R$
    \ELSE
        \STATE $(u_t, k_t) = \min_k \text{ in } R$
        \IF {$k_j > k_t$}
            \STATE Replace $(u_t, k_t)$ in $R$ with $(u_j, k_j)$
        \ENDIF
    \ENDIF
\ENDFOR
\RETURN $R$
\end{algorithmic}
\end{algorithm}

Our locality-aware sampling strategy leverages the weighted reservoir sampling algorithm as its core mechanism.
Given an initial node set $V_b=\{v_1,...v_i,...\}$, the graph sampling procedure performs neighbor sampling on each node $v_i$ according to predefined fanout settings.
In contrast to random sampling strategies, we adopt Algo.~\ref{bias-sample} to sample the neighborhood of $v_i$ to achieve a better cache hit rate.
Algo.~\ref{bias-sample} samples a subset $R$ of $m$ vertices from $\mathcal{N}(v_i)$, prioritizing vertices based on their weights $w$ which modulate sampling probability. 
The algorithm iteratively compares the weight  $k_j$ of each incoming vertex $u_j \in \mathcal{N}(v_i)$ to the minimum $k$ in $R$. 
If the replacement condition in line 7 is met, the vertex with $k_t$ is replaced by $u_j$.
Algo.~\ref{bias-sample} has a time complexity of $O(|\mathcal{N}(v_i)|)$ and a space complexity of $O(m)$.
In practice, setting higher weight values for cached vertices enables fine-grained control over preferential sampling of cached data.

Locality-aware sampling inherently prioritizes cached nodes. 
Consequently, the batch generation phase may eliminate a significant number of duplicate nodes, substantially alleviating GPU memory pressure.
While this shrinking of subgraphs inevitably introduces some accuracy degradation, experimental evidence across various datasets suggests that the robust graph topology can reduce this impact.
The relationship among graph density $d(G)$, locality-aware sampling (governed by the bias rate $\gamma$ and the feature cache volume $\Theta$), and accuracy drop $\Delta A$ can be modeled as Eq.~(\ref{eq:acc-drop}):
\begin{equation}
    \Delta A = f_1(\eta,\gamma,d(G),\Theta),
    \label{eq:acc-drop}
\end{equation}
where $\eta=\frac{|Vs_{i}|}{|V|}, Vs_i \in Gs_i(Vs_i, Es_i)$ is the overlapping ratio of subgraph for training and the entire graph.
Generally, increasing the bias rate $\gamma$ leads to greater accuracy drop; however, this effect is alleviated with growing cache volume $\Theta$.
Furthermore, graph partition before training can also reduce accuracy, because $\textrm {A}^{\textsubscript{3}}$GNN is then unable to access information from remote partitions without high-speed NVLink interconnects.
Based on this model, we can adaptively select sampling configurations. 
In the worst case, setting the bias rate to $\gamma=1$, effectively reverts the locality-aware strategy to naive random sampling, resulting in the same accuracy as existing works. 

In summary, locality-aware sampling offers a trade-off with reduced memory footprint in exchange for some drop in test accuracy, thereby contributing to affordable and adaptive GNN training, which can be validated with further ablation study.

\subsection{Multi-Level Parallelism Scheduling}

Optimizations targeting only the sampling stage offer limited gains beyond a certain point.
When sampling is no longer the training bottleneck, our focus should shift to optimizing the training stage, highlighting the need for efficient parallelization and pipelining.

Fixed parallelism scheduling has been widely used, yet different tasks benefit from different schemes.
For example, evaluations on Reddit dataset show that finer-grained parallelism increases throughput with more workers.
To address this, we designed an adaptive, multi-level parallelism scheduling.
It dynamically switches among pipeline strategies, balancing context switching overhead and throughput.

\begin{figure}[t]
    \centering
    \includegraphics[width=0.96\linewidth]{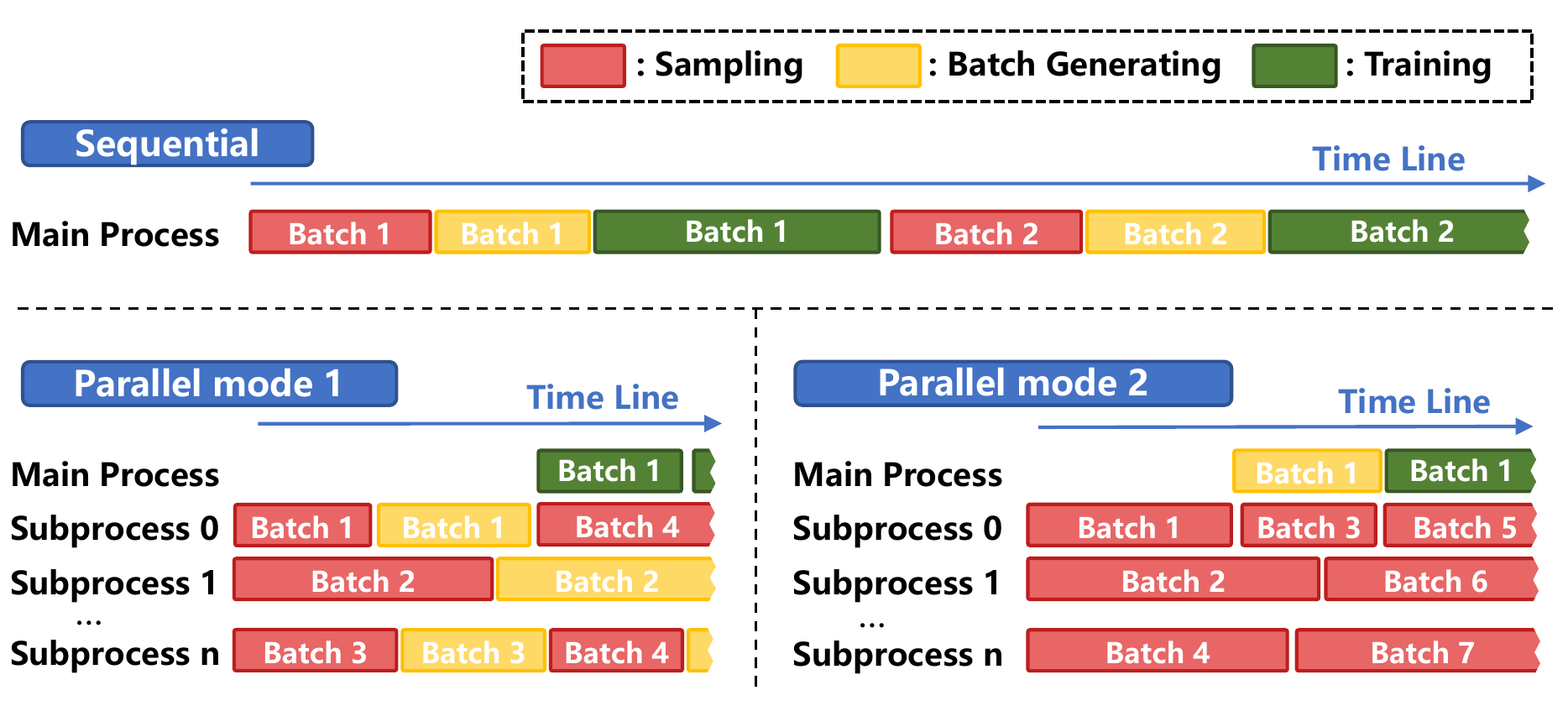}
    \vspace{-6pt}
    \caption{Parallelism scheduling templates of $\textrm {A}^{\textsubscript{3}}$GNN.}
    \label{fig:pipeline}
\end{figure}

Our multi-level parallelism has $3$ modes, as illustrated in Fig.~\ref{fig:pipeline}.
The sequential strategy minimizes overhead in constrained resources.
It executes each stage serially, eliminating data contention and reducing memory overhead.
However, the sequential execution also limits overall throughput. 
To address this limitation, parallel mode $1$ integrates and parallelizes sampling and batch generation.
According to Amdahl's Law, the throughput $\mathcal{T}_1$ of parallel mode $1$ can be expressed as Eq.~(\ref{eq:mode1-thr}), letting $n$ be the number of worker processes on each GPU:
\begin{equation}
    \mathcal{T}_1 = f_2(\max\{\frac{1}{n}(\mathcal{T}_{sample}+\mathcal{T}_{batch}),\mathcal{T}_{train}\} \cdot iter, d(G)).
    \label{eq:mode1-thr}
\end{equation}

However, mode $1$ increases GPU memory from process duplication, and introduces resource contention. 
Expected memory footprint $\mathcal{M}_1$ can be modeled as Eq.~(\ref{eq:mode1-mem}):
\begin{equation}
    \mathcal{M}_1=n \cdot f_3({num}_{GPU} \cdot \Theta, B, |M|, Runtime, d(G)).
    \label{eq:mode1-mem}
\end{equation}

The peak GPU memory footprint mainly consists of four components:
\begin{itemize}
    \item $B$: the mini-batch data after sampling, which is closely related to batch size and fanout settings.
    \item $|M|$: the GNN model parameters and intermediate results generated during forward and backward propagation.
    \item $Runtime$: the fixed memory footprint associated with the training stream context on GPUs.
    \item $\Theta$: the feature cache volume.
\end{itemize}

To accommodate a broader range of scenarios, we further incorporate an intermediate parallelism mode named parallel mode $2$, between the sequential mode and mode $1$.
Profiling based analysis reveals that resource contention arising from concurrent read/write operations is primarily concentrated within the batch generation stage.
Serializing this stage significantly reduces potential memory footprint, albeit at the cost of throughput drop.
The expected throughput and GPU memory footprint of mode $2$ can be formulated as Eq.~(\ref{eq:mode2-thr}) and Eq.~(\ref{eq:mode2-mem}):
\begin{equation}
    \mathcal{T}_2=f_4(\max\{\frac{1}{n}\mathcal{T}_{sample},\mathcal{T}_{batch}+\mathcal{T}_{train}\} \cdot iter, d(G)).
    \label{eq:mode2-thr}
\end{equation}
\begin{equation}
    \mathcal{M}_2=f_5({num}_{GPU} \cdot \Theta, B, |M|, n \cdot Runtime, d(G)).
    \label{eq:mode2-mem}
\end{equation}

Mode 2 mitigates the resource contention associated with concurrent access while simultaneously accelerating sampling through parallel execution. 
This mode yields a notable improvement in training throughput and maintains control over memory footprint.

In summary, we design a multi-level parallelism scheduling method that addresses the limitations of existing GNN training approaches.
Current methods rely on fixed parallelism schedules, limiting their adaptability across diverse workload scenarios and hardware configurations.
By analyzing the task scenarios and performance requirements, $\textrm {A}^{\textsubscript{3}}$GNN can tailor the parallel training strategy to address bottlenecks at different stages, which enables flexible trade-offs between GPU memory footprint and throughput.

\subsection{Task-Hardware Oriented Auto-Tuning}

Existing approaches often achieve satisfactory performance through manual tuning or default configurations, owing to their relatively simple design with fixed optimization strategies and limited reconfigurable parameters. 
However, $\textrm {A}^{\textsubscript{3}}$GNN introduces a more sophisticated framework that incorporates diverse sampling techniques and dynamic training scheduling strategies.
This expanded functionality results in an exponentially larger configuration space, rendering traditional manual tuning methods infeasible.
Thus, efficiently generating training configurations tailored to specific scenarios and task requirements becomes critical for further performance enhancement.

To enable task-hardware oriented auto-tuning, we propose a three-level mechanism comprising: 
\begin{itemize}
    \item Task-aware performance metrics prioritization,
    \item Hardware-aware constraints analysis,
    \item Multi-objective Pareto frontier exploration.
\end{itemize}

\begin{figure}[t]
    \centering
    \includegraphics[width=0.75\linewidth]{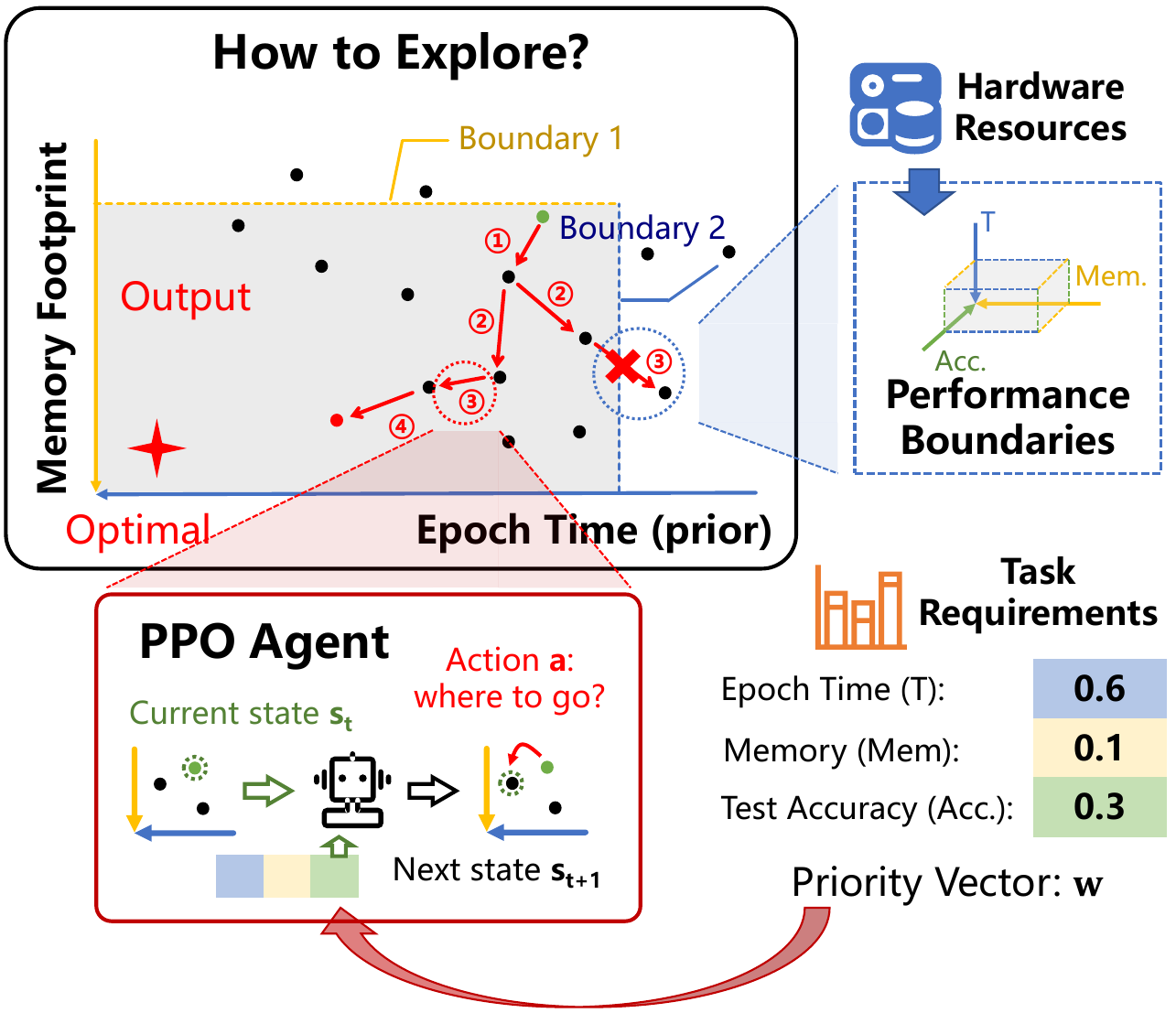}
    \vspace{-6pt}
    \caption{Task-hardware oriented auto-tuning.}
    \label{fig:autotuning}
\end{figure}

First, our \textbf{task-aware performance metrics prioritization} module dynamically adjusts evaluation criteria based on task-specific requirements.
As illustrated in Fig.~\ref{fig:autotuning}, we initialize the priority vector $\mathbf{w}$ with differentiated weights for various metrics (e.g., accuracy, throughput) according to their relative importance in the target training scenario.

For hardware considerations, the \textbf{hardware-aware constraints analysis} module formally maps physical resource constraints into optimization boundaries.
For instance, given a GPU memory capacity of $10$ GB, we enforce strict inequality constraints (e.g., peak memory consumption $< 10$ GB) during design space exploration (DSE).
When exploration approaches these boundaries, the weighted reward function will return extremely negative reward values $R$ to steer the search away from invalid configurations while maintaining optimization progress.

The core challenge then becomes efficiently predicting GNN training performance.
We extract the key configurations from  Eq.~(\ref{eq:acc-drop}-\ref{eq:mode2-mem}) and leverage existing open-source datasets and tasks to collect performance statistics for training a surrogate model. 
This surrogate model takes training configurations and graph characteristics as inputs and, by integrating sophisticated machine learning techniques such as XGBoost, Regression, and Decision Trees, generates accurate  performance predictions that guide the automated design space exploration.

Finally, our \textbf{multi-objective Pareto frontier exploration} employs the reinforcement learning-based DSE algorithm, which is detailed in Algo.~\ref{alg:ppo_gnn}.
We formulate the auto-tuning task as a Markov Decision Process (MDP) and employ the Proximal Policy Optimization (PPO) as reinforcement learning agent.
The state, denoted as $\mathbf{s}=\{\mathbf{p}, \mathbf{m}\}$, encompasses both the configuration settings and the corresponding predicted performance.
Here, $\mathbf{p}$ represents a specific configuration setting within the design space, and $\mathbf{m}=[thr, mem, acc]$ describes the performance metrics: throughput, memory footprint, and accuracy. 
The weighted reward formula (Line 10: $R_{t+1} \gets \mathbf{w}^\top \mathbf{m}_{t+1}$) inherently incorporates both task preferences and hardware constraints, enabling simultaneous optimization of competing objectives.
This formulation allows the discovery of Pareto-optimal configurations that optimally balance performance metrics within the feasible hardware constraints.

\begin{algorithm}[h]
\small % 使用更小的字体
\caption{PPO for $\textrm {A}^{\textsubscript{3}}$GNN Auto-Tuning}
\renewcommand{\algorithmicrequire}{\textbf{Input:}}
\renewcommand{\algorithmicensure}{\textbf{Output:}}
\label{alg:ppo_gnn}
\begin{algorithmic}[1]
\REQUIRE Metrics priority vector $\mathbf{w}$, initial configurations $\mathbf{p}_0$.
\ENSURE Recommend configurations $\mathbf{p}*$.
\WHILE{not converged}
    \STATE \textit{--- Parameter Adjustment ---}
    \STATE Get adjustment action: $\mathbf{a}_t \sim \pi_\theta(\mathbf{s}_t)$
    \STATE Update config.: $\mathbf{p}_{t+1} \gets \text{clip}(\mathbf{p}_t + \mathbf{a}_t, \text{valid\_range})$

    \STATE \textit{--- Termination Conditions ---}
    \STATE Predict performance with $\mathbf{p}_{t+1}$ $\rightarrow$ performance $\mathbf{m}_{t+1}$
    \IF{$\mathbf{m}_{t+1}$ violate constraints}
        \STATE $R_{t+1} \gets -\infty$
    \ELSE
        \STATE $R_{t+1} \gets \mathbf{w}^\top \mathbf{m}_{t+1}$
        \IF{$R_{t+1} > R^*$}
            \STATE $\mathbf{p}^* \gets \mathbf{p}_{t+1}$, $R^* \gets R_{t+1}$
        \ENDIF
    \ENDIF

    \STATE \textit{--- Policy Update ---}
    \STATE Store $(\mathbf{s}_t, \mathbf{a}_t, R_{t+1}, \mathbf{s}_{t+1})$ in buffer
    \STATE Update $\pi_\theta$ using clipped objective $\mathcal{L}_{\text{clip}}$
    \STATE Update $V_\phi$ via TD-learning
    
\ENDWHILE
\end{algorithmic}
\end{algorithm}

This three-level auto-tuning mechanism enables the automated generation of adaptive training parameter configurations for diverse scenarios.
Furthermore, task-hardware oriented auto-tuning is built upon the preceding adaptive training strategies.
It further fine-tunes GNNs training performance from a fine-grained perspective, resulting in superior overall performance.
Collectively, these innovations detailed in Sec.~\ref{sec:framework} establish a GNN training framework optimized for affordability, adaptivity, and automation.

\section{Experiment}
\label{sec:exp}

\subsection{Experimental Settings}

\textbf{Baselines and datasets.}
To evaluate $\textrm {A}^{\textsubscript{3}}$GNN, we employ $2$ baselines: the state-of-the-art GNN training framework Quiver~\cite{quiver2023}, and one of the most widely used PyTorch Geometric (PyG) framework~\cite{Fey/Lenssen/2019}.
Our experiments are conducted on the publicly available OGB datasets~\cite{hu2020ogb} for node classification tasks.
In addition, all experimental and profiling results are obtained using NVIDIA tools, taking the average results of $10$ runs.
For a fairer comparison with manual optimizations in Quiver and PyG~\cite{quiver2023,Fey/Lenssen/2019}, we deploy them on our compute platforms, and measure their performance.

\begin{table}[h]
    \renewcommand\arraystretch{1.1}
    \centering
    \caption{Design Space Definition.}
    \vspace{-8pt}
    \begin{tabular}{|c|c|c|}
    \hline
        & \textbf{Configurations} & \textbf{Range} \\ \hline
        \multirow{2}{*}{\textbf{General}} & Batch Size & [64, 1024] \\ \cline{2-3}
        & Graph Partition & [1,u] \\ \hline
        \multirow{3}{*}{\textbf{Sampling}} & Bias Rate & [1,$\infty$) \\ \cline{2-3}
         & Sampling Device & \{CPU,GPU\} \\ \cline{2-3}
         & Workers & [1,$\infty$) \\ \hline
        \textbf{Feature} & Cache Volume & [0,$\infty$)  \\ \hline
        \textbf{Parallelism} & Mode & \{Sequential, P-mode 1, P-mode 2\} \\
        \hline
    \end{tabular}
    \label{tab:designspace}
    \vspace{-3pt}
\end{table}

\textbf{$\textrm {A}^{\textsubscript{3}}$GNN Settings.}
$\textrm {A}^{\textsubscript{3}}$GNN defines a set of reconfigurable settings to cover its large configuration space. 
As illustrated in Tab.~\ref{tab:designspace}, $\textrm {A}^{\textsubscript{3}}$GNN's training performance can be finely tuned by combining diverse strategies and configuring their settings to align with specific task requirements.
The changeable configurations are categorized into four classes: 
1) general configurations, such as batch size and the number of subgraphs to partition; 
2) sampling-oriented configurations, including the bias rate for locality-aware sampling, the execution platform for sampling (GPU or CPU), and the number of worker processes; 
3) graph feature-related configurations, such as cache volume, representing the amount of GPU memory allocated for caching; 
and 4) parallelism modes, for which we design three distinct modes – sequential, parallel mode 1, and parallel mode 2 – each exhibiting different performance characteristics.

\textbf{Compute Platforms.}
We employ 4 different compute platforms for comparing $\textrm {A}^{\textsubscript{3}}$GNN and competitors: 1) A100-Server equipped with two NVIDIA A100(80G) and Intel 6330 CPU, 2) 2080-Server equipped with seven RTX2080Ti and Intel 5118 CPU, 3) 4090-Workstation equipped with RTX4090 and Intel i9-13900K, 4) M40-Server equipped with two Tesla M40 and Intel E5-2620 CPU.

\subsection{$\textrm {A}^{\textsubscript{3}}$GNN over Existing GNNs Training Frameworks}

\begin{table}[t]
    \renewcommand\arraystretch{1.1}
    \centering
    \caption{GNN Training Performance Comparison.}
    \vspace{-8pt}
    \begin{tabular}{|c|c|c|c|c|c|}
    \hline
        \textbf{Platforms}  & \textbf{Datasets} & \textbf{Frame.} & \makecell[c]{\textbf{Thr.}\\ \textbf{[ep./s]}} & \makecell[c]{\textbf{Mem.} \\ \textbf{[GB]}} & \textbf{Acc.} \\ \hline
        \multirow{8}{*}{
            \makecell[c]{GPU: \\ 2*A100,\\
                         CPU:\\6330}
                    } & \multirow{4}{*}{Reddit} 
           & PyG(new) & 0.336 & 18.492  & 95.15\% \\ 
         & & Quiver   & 0.217 & 16.587  & 95.28\% \\ 
         & & Ours(T*) & \textbf{0.838} & 136.644 & 95.06\% \\ 
         & & Ours(M*) & 0.052 & \textbf{11.324}  & 95.05\% \\ \cline{2-6} 
         & \multirow{4}{*}{Products} 
           & PyG(new) & 0.244 & 18.777  & 74.13\% \\  % Namespace(dataset='ogbn-products', hop_1_neighbor=25, hop_2_neighbor=15, hidden_channels=128, backend='PYG', partition='cluster', part_num=1, sample_dev='CPU', cache_volume='1G', bias_rate=1.0, epochs=5, batch_size=1024, num_workers=4, persistent_workers=1, filter_per_worker=1, infer_flag=0)
         & & Quiver   & 0.237 & 11.714  & 74.86\% \\ 
         & & Ours(T*) & \textbf{0.884}  & 102.691 & 75.62\% \\  
         & & Ours(M*) & 0.078 & \textbf{5.664} & 75.97\% \\ \hline % 64进程
        \multirow{8}{*}{
            \makecell[c]{GPU:\\7*2080Ti,\\
                         CPU:\\5118}
                    } & \multirow{4}{*}{Reddit} 
           & PyG(new) & 0.151 & 38.888 & 94.97\% \\ 
         & & Quiver   & 0.189 & 43.912 & 94.89\% \\ 
         & & Ours(T*) & \textbf{0.317} & 48.910 & 94.96\% \\ 
         & & Ours(M*) & 0.077 & \textbf{17.978} & 94.99\% \\ \cline{2-6} 
         & \multirow{4}{*}{Products} 
           & PyG(new) & 0.112 & 37.447 & 72.19\% \\ 
         & & Quiver   & 0.144 & 20.630 & 73.87\% \\ 
         & & Ours(T*) & \textbf{0.443} & 31.125 & 74.54\% \\ 
         & & Ours(M*) & 0.219 & \textbf{6.132}  & 75.07\% \\ \hline
        \multirow{8}{*}{
            \makecell[c]{GPU:\\4090, \\
                         CPU:\\i9-13900K}
                    } & \multirow{4}{*}{Reddit} 
           & PyG(new) & 0.350 & 8.920 & 94.94\% \\ 
         & & Quiver   & 0.205 & 8.309 & 95.02\% \\ 
         & & Ours(T*) & \textbf{0.447} & 23.63 & 94.95\% \\ 
         & & Ours(M*) & 0.044 & \textbf{5.182} & 95.04\% \\ \cline{2-6}
         & \multirow{4}{*}{Products} 
           & PyG(new) & 0.221 & 7.980 & 74.89\% \\ 
         & & Quiver   & 0.227 & 4.758 & 75.52\% \\ 
         & & Ours(T*) & \textbf{0.629} & 23.55 & 74.07\% \\ 
         & & Ours(M*) & 0.044 & \textbf{2.096} & 74.87\% \\ \hline
        \multirow{8}{*}{
            \makecell[c]{GPU:\\2*M40, \\
                         CPU:\\E5-2620}
                    } & \multirow{4}{*}{Reddit} 
           & PyG(new) & 0.061 & 12.498 & 95.04\% \\ 
         & & Quiver   & 0.070 & 12.074 & 95.57\% \\ 
         & & Ours(T*) & \textbf{0.138} & 21.617 & 95.14\% \\ 
         & & Ours(M*) & 0.022 & \textbf{9.820}  & 94.79\% \\ \cline{2-6}
         & \multirow{4}{*}{Products} 
           & PyG(new) & 0.061 & 7.902  & 74.64\% \\ 
         & & Quiver   & 0.071 & 6.177  & 74.72\% \\ 
         & & Ours(T*) & \textbf{0.180} & 11.949 & 76.50\% \\ 
         & & Ours(M*) & 0.037 & \textbf{1.734} & 75.10\% \\ 
    \hline
    \end{tabular}
    \label{tab:overall}
    \vspace{-3pt}
\end{table}

We evaluate GNN training performance against state-of-the-art approaches across diverse training tasks.
As summarized in Tab.~\ref{tab:overall}, GNN training tasks were compared on platforms ranging from server-class systems (A100 \& Intel 6330) to desktop workstations (e.g., a 4090).
Owing to the extensive volume of experimental data, Tab.~\ref{tab:overall} presents the performance results obtained on two commonly used datasets only.
In the table, we evaluate $3$ performance metrics, \textbf{Thr.[ep./s]} represents throughput (epochs per second), \textbf{Mem.[GB]} denotes peak GPU memory footprint, and \textbf{Acc.} indicates test accuracy.
Results with \textbf{T*} reflect a solution of $\textrm {A}^{\textsubscript{3}}$GNN that prioritizes throughput, allowing for some compromise in GPU memory footprint and accuracy.
Our approach significantly improves training speed, attributed to the integration of locality-aware sampling and multi-level parallelism scheduling, achieving speedups of up to $3.43\times$, $3.95\times$, $2.84\times$, $2.95\times$ on different platforms, albeit with increased memory footprint.
Results with \textbf{M*} mainly target resource constrained platforms.
It minimizes GPU memory footprint by combining sequential training execution and data locality enhancement.
Importantly, $\textrm {A}^{\textsubscript{3}}$GNN enables not only the prioritization of a single performance metric at the expense of others, but also balanced performance.
Note that we get the \textbf{T*}, and \textbf{M*} from the two ends of the Pareto front.
There are many other choices on Pareto front if we make some trade-off among different metrics.
For instance, a balanced configuration of $\textrm {A}^{\textsubscript{3}}$GNN on A100 GPUs, sacrificing some acceleration for improved memory efficiency and accuracy, yields a $1.88\times$ acceleration, with only $35\%$ overhead in GPU memory footprint, and negligible accuracy drop.

\begin{figure}[t]
    \centering
    \includegraphics[width=1\linewidth]{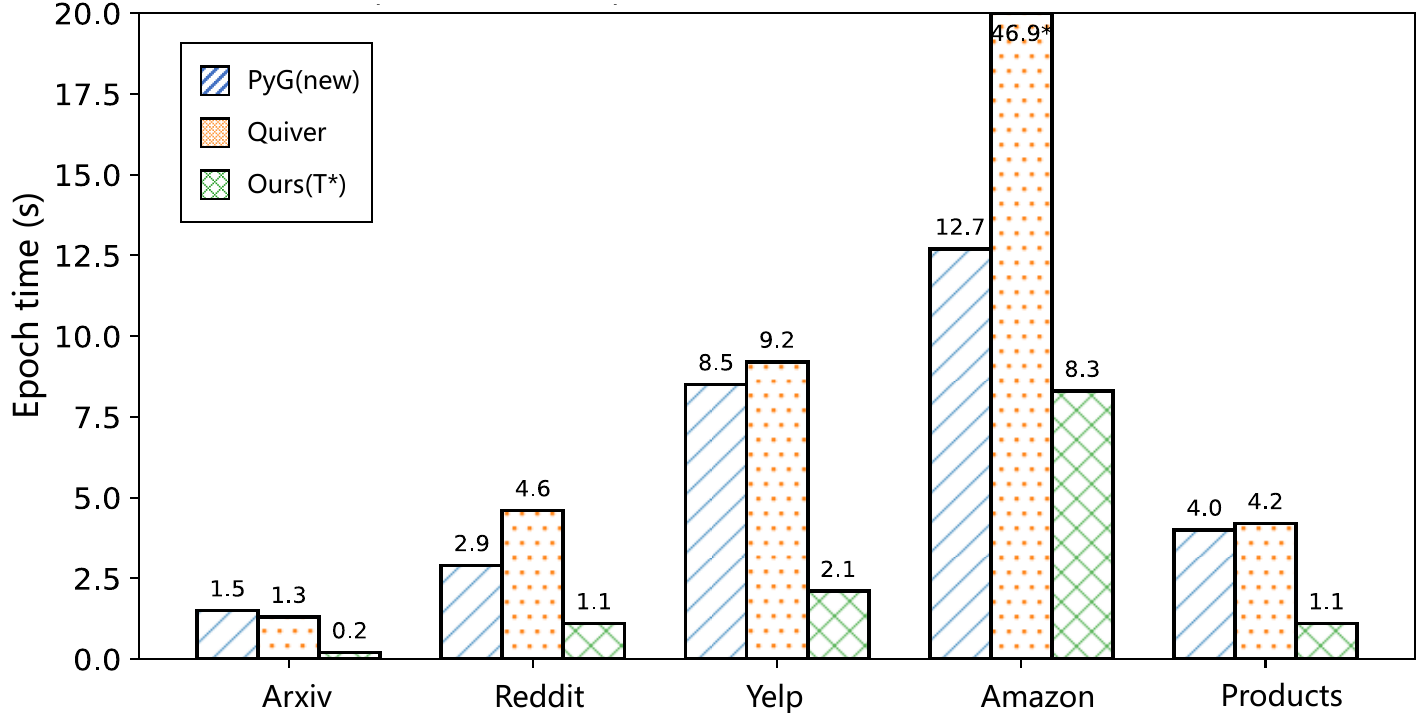}
    \vspace{-18pt}
    \caption{Scalability of $\textrm {A}^{\textsubscript{3}}$GNN across different datasets.}
    \label{fig:scalableds}
    
\end{figure}

Furthermore, $\textrm {A}^{\textsubscript{3}}$GNN exhibits broad applicability across diverse datasets. 
Benchmarking experiments conducted on A100 GPUs demonstrate that $\textrm {A}^{\textsubscript{3}}$GNN consistently achieves significant performance gains on datasets of varying scales. 
As shown in Fig.~\ref{fig:scalableds}, we conduct evaluations on more datasets with different characteristics, including lightweight graph such as Arxiv ($169,343$ nodes \& $1,166,243$ edges), sparse but large-scale graphs such as Products ($2,449,029$ nodes \& $61,859,140$ edges), and dense graphs like Amazon ($1,569,960$ nodes \& $264,339,468$ edges), Yelp and Reddit.
$\textrm {A}^{\textsubscript{3}}$GNN achieves an average of $3.88\times$ acceleration across these $5$ datasets.

\subsection{Ablation Studies}
To dissect the contributions of individual strategies, we conduct controlled experiments to analyze the impact of locality-aware sampling, multi-level parallelism, and other strategies on training performance. 
This analysis further illuminates how these strategies collectively enable affordable, adaptive, and automatic GNN training.

\textbf{Impact of Locality-aware Graph Sampling.} 
To facilitate controlled experimental analysis, we employed the simplest sequential mode to minimize interference from other strategies. 
For the feature cache, we adopt a static caching policy, mirroring the approach in prior work such as Quiver~\cite{quiver2023} and DSP~\cite{cai2023dsp}. 
We cached the most frequently accessed nodes, with a fixed cache volume of $40$MB.

\begin{figure}[t]
    \centering
    \includegraphics[width=1\linewidth]{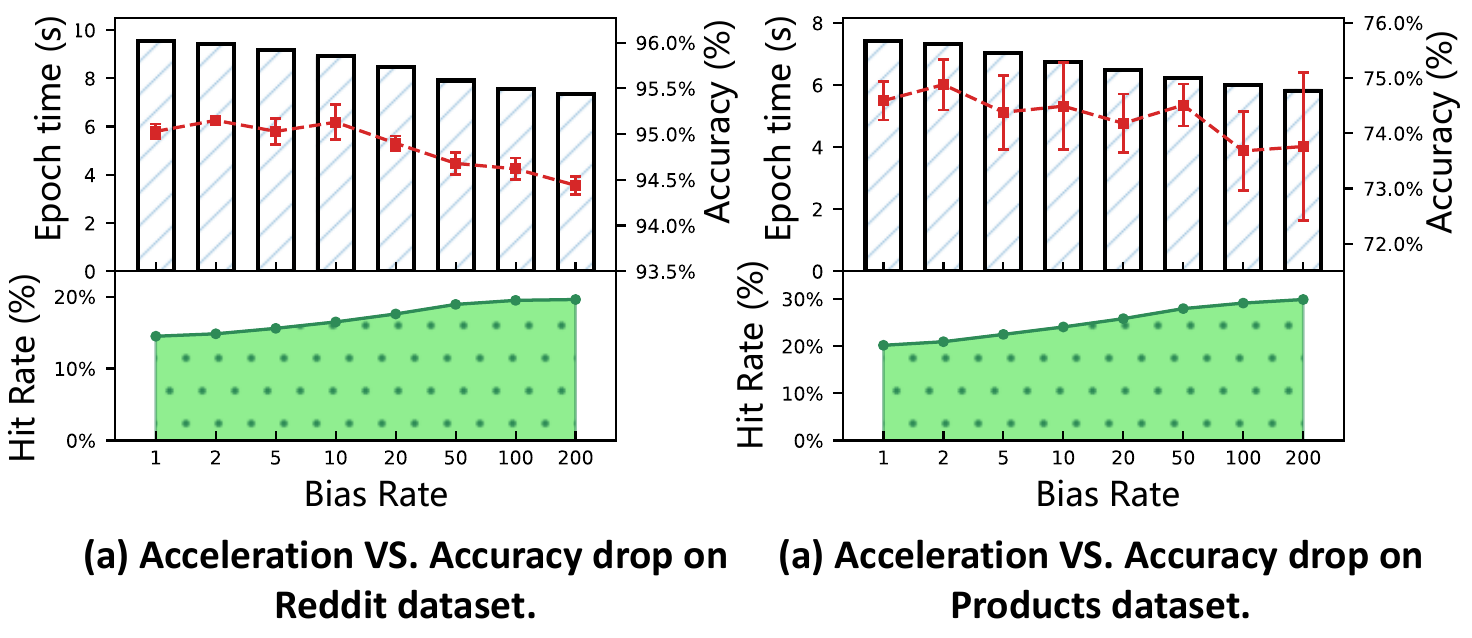}
    \vspace{-18pt}
    \caption{Impact of locality-aware graph sampling strategies. Bars show average epoch time, red lines represent model accuracy on the test set, and green lines illustrate cache hit rate variations with bias rate.}
    \label{fig:bias-sample}
\end{figure}

To evaluate the impact of locality-aware sampling, we conduct ablation studies on both the Products and Reddit datasets.
As illustrated in Figure~\ref{fig:bias-sample}, we adjust the bias rate $\gamma$ of the locality-aware sampling to prioritize the selection of cached data, resulting in a substantial increase in the cache hit rate. Yielding a $30$\% and $27$\% enhancement in throughput on the Reddit and Products datasets, respectively, at the cost of a $1$\% accuracy drop.
 
\begin{figure}[h]
    \centering
    \includegraphics[width=1\linewidth]{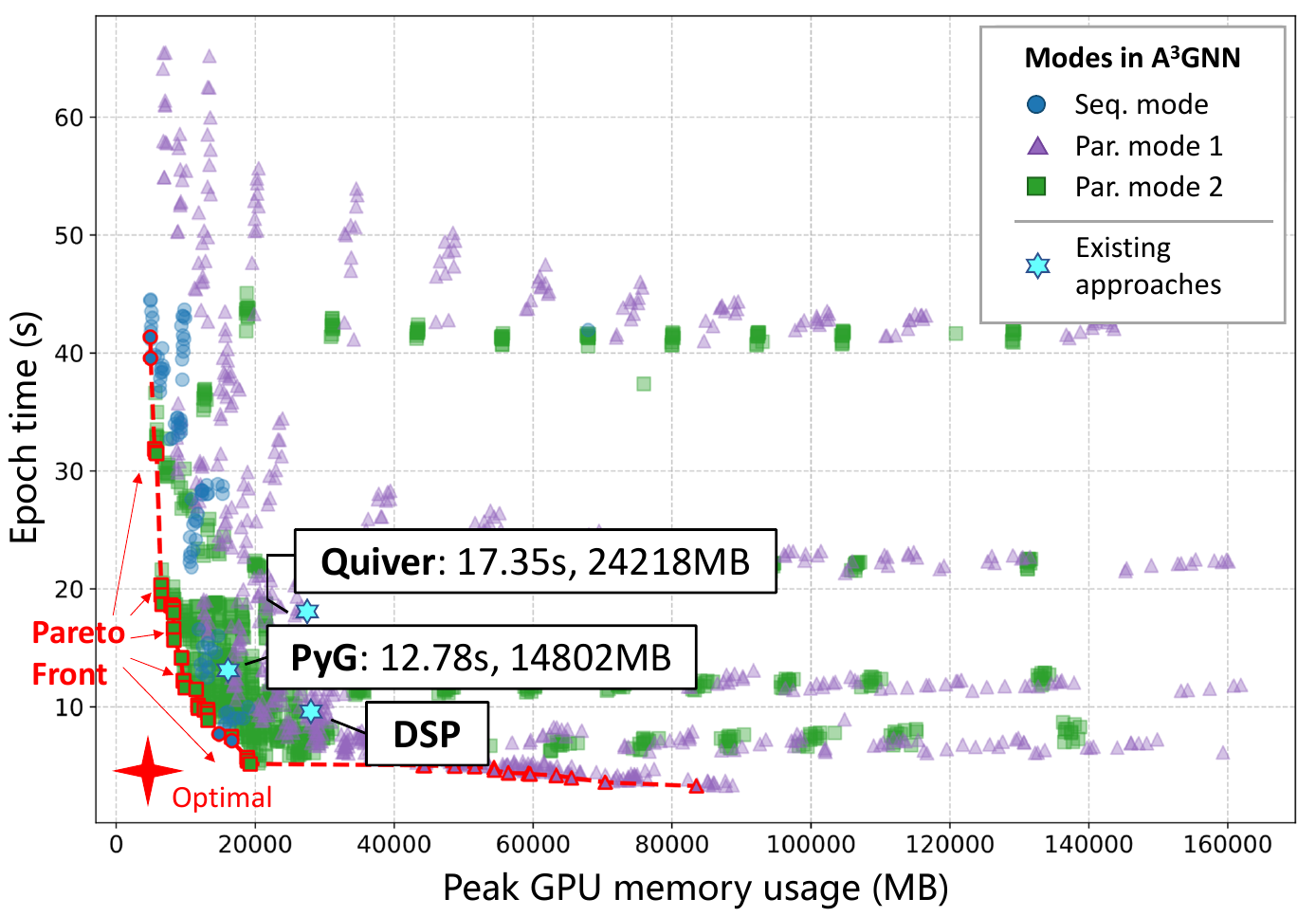}
    \vspace{-20pt}
    \caption{Performance analysis of multi-level parallelism scheduling. Each vertex represents the ground-truth performance of $\textrm {A}^{\textsubscript{3}}$GNN under different parallelism scheduling modes (indicated by colors). Red-lined vertices form the Pareto front, and existing works' performance is marked for comparison.}
    \label{fig:ml-para}
\end{figure}

\textbf{Impact of Multi-Level Parallelism Scheduling.}
We assess the effectiveness of our proposed multi-level parallelism scheduling modes on the Reddit dataset.
We benchmark the training performance of sequential mode and parallel modes $1$ and $2$, we consider throughput and GPU memory footprint as illustrative examples of the inherent trade-offs.
To ensure a fair comparison, we adopted the default settings from relevant open-source works and compared the results with our findings.
$\textrm {A}^{\textsubscript{3}}$GNN achieves a 3.3× higher throughput than Quiver and a 1.8× higher throughput than PyG, while maintaining comparable memory consumption. 
Since the source codes for certain prior works are not available, we estimate their performance based on comparisons reported in their respective publications.
As depicted in Fig.~\ref{fig:ml-para}, we present the training performance for all parameter settings as a scatter plot, with Pareto front illustrating the performance envelope of each mode.

These results demonstrate that $\textrm {A}^{\textsubscript{3}}$GNN can adaptively achieve strong experimental results across diverse scenarios by selecting appropriate training configurations.
Examining the Pareto frontier reveals that each mode is represented on the boundary under varying demands.
Empirically, the sequential mode, with its economical resource consumption, performs better when GPU memory is limited.
Parallel mode $2$ is most effective when balancing resource usage and performance under resource constraints.
Finally, parallel mode $1$ is optimal in scenarios with ample GPU resources, enabling maximum throughput.
The adaptive mode selection further highlights the importance of adaptive training.

\textbf{Impact of Task-Hardware Oriented Auto-Tuning.}
The main idea of task-hardware oriented auto-tuning is to design a surrogate model to predict GNNs training performance.
It employs a reinforcement learning-based agent to determine the training configuration settings.
To illustrate the effect of our auto-tuning approach, we conduct ablation studies on the accuracy of surrogate model on performance predictions, the overhead of auto-tuning, and the performance comparison between output settings of the PPO agent and the ground truth.

We benchmark the performance prediction and ground-truth profiling results on Products, Yelp and reddit datasets.
\begin{table}[h]
    \renewcommand\arraystretch{1.1}
    \centering
    \caption{$R^2$ Score of Performance Prediction.}
    \vspace{-8pt}
    \begin{tabular}{|c|c|c|c|}
    \hline
        \textbf{Performance Metrics} & \textbf{Reddit} & \textbf{Yelp} & \textbf{Ogbn-products} \\ \hline
        Throughput & 0.8361 & 0.7921 & 0.7303          \\ \hline
        Memory Footprint & 0.8531 & 0.8841 & 0.7264   \\
        \hline
    \end{tabular}
    \label{tab:correctness}
    \vspace{-3pt}
\end{table}

Note that \textit{$R^2$ scores} indicate better precision of estimators when they are closer to $1.0$.
As shown in Tab.~\ref{tab:correctness}, our surrogate model can make relative accurate prediction.
Furthermore, the PPO agent can nearly achieve optimal performance, and the exploration is on average $2.1\times$ faster than grid search exploration.

\section{Conclusion}

In this work, we present $\textrm {A}^{\textsubscript{3}}$GNN, a novel GNN training framework designed for Affordable, Adaptive, and Automatic GNN training.
Leveraging locality-aware sampling, $\textrm {A}^{\textsubscript{3}}$GNN attains high cache hit rates while minimizing resource overhead.
This optimization yields substantial improvements in both memory footprint and throughput, making efficient GNN training accessible even with less powerful devices.
The proposed multi-level parallelism scheduling mechanism enables adaptive selection of appropriate parallel strategies based on scenario characteristics, leading to significant improvements in accuracy, memory footprint, and throughput, respectively.
To further enhance performance, $\textrm {A}^{\textsubscript{3}}$GNN employs automated tuning to determine optimal configuration settings that satisfy hardware constraints while maximizing resource utilization.
Comprehensive experiments on diverse tasks and platforms show that $\textrm {A}^{\textsubscript{3}}$GNN delivers $1.88\times$ speedup in GNN training with minimal accuracy degradation and only $35.1\%$ additional memory consumption compared to state-of-the-art methods.
With flexible performance trade-offs, $\textrm {A}^{\textsubscript{3}}$GNN can further achieve up to $3.95\times$ acceleration or reduce GPU memory usage to just $16.3\%$ of baseline approaches.

\bibliographystyle{IEEEtran}
\small
\bibliography{iccad2025}

\end{document}